\documentclass[a4paper]{article}
\usepackage{Odyssey2020}
\usepackage{epsfig,amsmath,graphicx,amssymb,xcolor,color,dsfont}
\usepackage{booktabs}
\usepackage{multirow}

\usepackage[hidelinks]{hyperref}
\usepackage{booktabs, float, verbatim}
\usepackage{cite}

\newcommand{\Mu}{\boldsymbol{\mu}}
\newcommand{\x}{\boldsymbol{x}}
\newcommand{\transpose}{^{^{\intercal}}}
\newcommand{\inv}{^{^{-1}}}

\usepackage{booktabs}
\usepackage{multirow}
\usepackage{tabularx}
\newcolumntype{Y}{>{\centering\arraybackslash}X}
\usepackage{lipsum}

\newcommand\blfootnote[1]{%
  \begingroup
  \renewcommand\thefootnote{}\footnote{#1}%
  \addtocounter{footnote}{-1}%
  \endgroup
}
\title{NPLDA: A Deep Neural  PLDA Model for Speaker Verification}

\name{{Shreyas Ramoji, Prashant Krishnan, Sriram Ganapathy}}

\address{Learning and Extraction of Acoustic Patterns (LEAP) Lab, Department of Electrical Engineering\\Indian Institute of Science, Bengaluru, India\\
{\small \tt \{shreyasr, prashantkv1, sriramg\}@iisc.ac.in}}
\begin{document}
\maketitle

\begin{abstract}
The state-of-art approach for  speaker verification  consists of a neural network based embedding extractor along with a backend generative model such as the Probabilistic Linear Discriminant Analysis (PLDA). 
In this work, we propose a neural network approach  for backend modeling in speaker recognition. The likelihood ratio score of the generative PLDA model is posed as a discriminative similarity function and the learnable parameters of the score function are optimized using a verification cost. The proposed model, termed as neural PLDA (NPLDA), is initialized using the generative PLDA model parameters. The loss function for the NPLDA model is an approximation of the minimum detection cost function (DCF). The speaker recognition experiments using the NPLDA model are performed on the speaker verificiation task in the VOiCES datasets as well as the SITW challenge dataset. In these experiments, the NPLDA model optimized using the proposed loss function improves significantly over the state-of-art PLDA based speaker verification system.
\end{abstract}

\section{Introduction}\label{sec:Intro}
One of the earliest successful approach to speaker recognition used the Gaussian mixture modeling (GMM) from the training data followed by an adaptation using maximum-aposteriori (MAP) rule \cite{reynolds2000speaker}. The development of i-vectors as fixed dimensional front-end features for speaker recognition tasks was introduced in \cite{kenny2007joint,dehak2011front}.  In the recent years, neural network embeddings trained on a  speaker discrimination task were proposed as features to replace the i-vectors. These features called x-vectors \cite{snyder2018x} were shown to improve over the i-vectors for speaker recognition~\cite{mclaren2018train}. 

Following the extraction of x-vectors/i-vectors, different pre-processing steps are employed to transform the embeddings. The common steps include linear discriminant analysis (LDA)~\cite{dehak2011front}, unit length normalization~\cite{garcia2011analysis} and within-class covariance normalization (WCCN)~\cite{hatch2006within}. The transformed vectors are modeled with probabilistic linear discriminant analysis (PLDA)~\cite{kenny2010bayesian}. The PLDA model is used to compute a log likelihood ratio from a pair of enrollment and test embeddings which is used to verify whether the given trial is a target or non-target. 

In this paper, we propose a neural backend model which jointly performs pre-processing and scoring. This model, refered to as neural PLDA (NPLDA), operates on pairs of x-vector embeddings (a pair of enrollment and test x-vectors), and outputs a score that allows the decision of target versus non-target hypotheses. 
The implementation using neural layers allows the entire model to be learnt using a speaker verification cost. The use of conventional cost functions like binary cross entropy cause overfitting of the model to the training speakers, and have generalization issues on evaluation sets. In an attempt to avoid this, we use an approximation to the minimum detection cost (minDCF) \cite{van2007introduction} to optimize the neural backend model. With several experiments on speakers in the wild (SITW) dataset and the VOiCES development and evaluation datasets, we show that the proposed approach improves significantly over the state-of-the-art x-vector based PLDA system. 



The rest of the paper is organized as follows. In Section~\ref{sec:related_work}, we highlight relevant prior work done in the field of discriminative backend for speaker verification.  The datasets used in training and testing are described in Section~\ref{sec:data}. Section \ref{sec:fe_feature_xvector} describes the front-end configurations used for feature processing and x-vector extraction. The past approaches to backend modeling in speaker verification are discussed in Section~\ref{sec:prevBackend}. Section \ref{sec:PldaNet} describes the proposed neural network architecture used, and the connection with generative PLDA model. In Section \ref{sec:costfuncs}, we present a smooth approximation to the detection cost function which is used as an objective to optimize in a neural network.  This is followed by discussion of results in Section \ref{sec:results} and a brief set of concluding remarks in Section \ref{sec:summary}.

\section{Related Prior Work}
\label{sec:related_work}
 The common approaches for scoring in speaker verification systems  include support vector machines (SVMs) \cite{campbell2006support}, Gaussian backend model \cite{mclaren2013adaptive,benzeghiba2009language} and the probabilistic linear discriminant analysis (PLDA) \cite{kenny2010bayesian}. Some efforts on pairwise generative and discriminative modeling are discussed in \cite{cumani2013pairwise,cumani2014large,cumani2014generative}. The discriminative version of PLDA with logistic regression and support vector machine (SVM) kernels has also been explored in ~\cite{burget2011discriminatively}. In this work, the authors use the functional form of the generative model and pool all the parameters needed to be trained into a single long vector. These parameters are then discriminatively trained using the SVM loss function with pairs of input vectors. The discriminative PLDA (DPLDA) is however prone to over-fitting on the training speakers and leads to degradation on unseen speakers in SRE evaluations~\cite{villalba2019state}. The regularization of embedding extractor network using a Gaussian backend scoring  has been investigated in \cite{ferrer2019optimizing}. 
 
 Recently, end-to-end approaches to speaker verification have also been examined. For example, in~\cite{rohdin2018end}, the i-vector extraction with PLDA scoring has been jointly derived using a deep neural network architecture and the entire model is trained using a binary cross entropy training criterion. The use of triplet loss in end-to-end speaker recognition has shown promise for short utterances~\cite{zhang2017end}.  Wan et. al.~\cite{wan2018generalized} proposed a generalized end-to-end loss inspired by minimizing the centroid mean of within speaker distances while maximizing across speaker distances. However, in spite of these efforts, most of the successful systems for SRE evaluations continue to use the generative PLDA backend model. 
 
In this paper, we argue that the major issue of over-fitting in discriminative backend systems arises from the choice of the model and the loss function. In the detection cost function for the VOiCES task, the false-alarm errors have more significance compared to miss errors. Thus, incorporating the DCF metric directly in the optimization should aid the verification task. Recent developments in this direction in this direction includes efforts in using the approximate DCF metric for text dependent speaker verification \cite{Mingote2019}. Further, by training multiple pre-processing steps along with the scoring module, the model learns to generate representations that are better optimized for the speaker verification task. 

\section{Dataset Description}\label{sec:data}
The VOiCES far field speaker recognition challenge \cite{Nandwana2019} had two training conditions - \emph{fixed} and \emph{open}. This paper presents the models trained only for the fixed training condition.  

\subsection{Training Data}
The models reported in this work (front-end and backend models) are trained entirely using speech data extracted from combined VoxCeleb 1 and 2 corpora~\cite{nagrani2017voxceleb, chungvoxceleb2}. These datasets contain speech extracted from celebrity interview videos available on YouTube, spanning a wide range of different ethnicity, accents, professions, and ages. For training the x-vector extractor, we use about $1.2$M segments from $7323$ speakers selected from VoxCeleb 1 (dev and test), and VoxCeleb 2 (dev). 

\subsection {Development and Test Data}
We test the models using VOiCES dataset~\cite{nandwana2019voices} and the  speakers in the wild (SITW) \cite{McLaren+2016} datasets.

The SITW dataset consists of nearly $300$ spakers across clean interview, red carpet interviews, stadium conditions, outdoor conditions. The \emph{core} (single speaker) enrollment and test sets consists of around $6000$ segments and the \emph{core-core} test  condition consists of around $800,000$ trials.

The VOiCES development set was created by partioning the audio files from Rooms 1 and 2 (out of the four available) of the VOiCES corpus~\cite{nandwana2019voices}. The development set consisted of $15,904$ audio segments from $196$ speakers. Each audio recording contains only one speaker.  The development set represented different rooms, microphones, noise distractors, and loudspeaker angles.  The development set included $20,224$ target and $4,018,432$ impostor trials.  The evaluation set was created by partioning  the  audio  recordings  of  the  VOiCES  corpus  from  Rooms  3  and 4.  It consisted of $11,392$ audio segments  from  $100$  speakers  that  were  disjoint  from  the  development set.  The evaluation set included $36,443$ target and $357,073$ impostor trials.

\section{Front-end Model - X-vector Extractor}\label{sec:fe_feature_xvector}
In this section, we provide the description of the front-end feature extraction and x-vector model configuration.
\subsection{Training }
The x-vector extractor is trained entirely using speech data extracted from combined VoxCeleb 1 ~\cite{nagrani2017voxceleb} and VoxCeleb 2 corpora~\cite{chungvoxceleb2}. These datasets contain speech extracted from celebrity interview videos available on YouTube, spanning a wide range of different ethnicities, accents, professions, and ages. For training the x-vector extractor, we use $1,276,888$ segments from $7323$ speakers selected from Vox-Celeb 1 (dev and test), and VoxCeleb 2 (dev).

This x-vector extractor was trained using $30$ dimensional Mel-Frequency Cepstral Coefficients (MFCCs) from $25$ ms frames shifted every $10$ ms using a $23$-channel mel-scale filterbank spanning the frequency range $20$ Hz - $7600$ Hz. A 5-fold augmentation strategy is used that adds four corrupted copies of the original recordings to the training list~\cite{snyder2018x,mclaren2018train}. 
The augmentation step  generates
$6,384,440$ training segments for the combined VoxCeleb set.

An extended TDNN with $12$ hidden layers and rectified linear unit (RELU) non-linearities is trained to discriminate among the nearly $7000$ speakers in the training set~\cite{mclaren2018train}. The first $10$ hidden layers operate at frame-level, while the last $2$ layers operate at segment-level.   There is a $1500$-dimensional statistics pooling layer between the frame-level and segment-level layers that accumulates all frame-level  outputs using mean and standard deviation.  After training, embeddings are extracted from the $512$ dimensional affine component of the $11$th layer (i.e., the first segment-level layer).  More details regarding the DNN architecture and the training process can be found in \cite{mclaren2018train}.

 \begin{figure*}[t!]
 \begin{center}
    \includegraphics[width=0.79\linewidth, trim={2.2cm 2.1cm 1.5cm 1.5cm},clip]{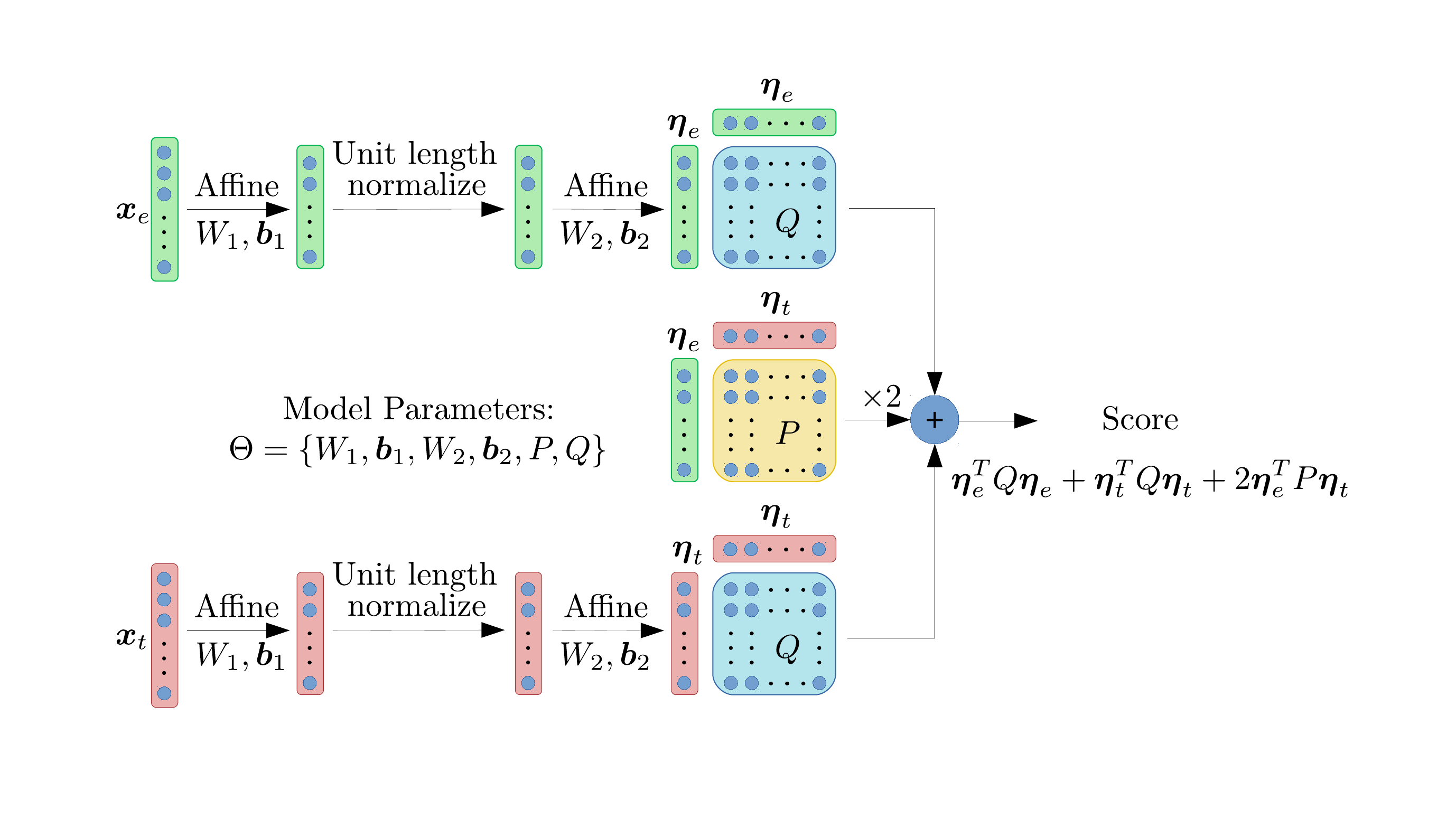}
    \caption{Neural PLDA Net Architecture: The two inputs $\x_e$ and $\x_t$ are the enrollment and test x-vectors respectively.}
    \label{fig:PldaNet}
    \vspace{-3ex}
\end{center}
\end{figure*}

\section{Approaches to backend Modeling}\label{sec:prevBackend}
In this section, we describe the mathematical details of the past methods used for backend modeling in speaker verification. These include the generative PLDA modeling~\cite{kenny2010bayesian,sizov2014unifying}, discriminative PLDA modeling proposed in a support vector framework~\cite{burget2011discriminatively} and the Gaussian backend model~\cite{cumani2013pairwise}. 

\subsection{Generative Gaussian PLDA (GPLDA) }
Following the x-vector extraction, the embeddings are centered (mean removed), transformed using LDA and unit length normalized. The PLDA model on the processed x-vector for a given recording  is,
\begin{equation}
{\boldsymbol \eta} _r = \Phi \boldsymbol{\omega} + \boldsymbol{\epsilon}_r 
\end{equation}
where ${\boldsymbol \eta} _r$ is the x-vector for the given recording, $\boldsymbol{\omega}$ is the latent speaker factor with a Gaussian     prior of $\mathcal{N}(\textbf{0},\textbf{I})$, $\Phi$ characterizes the speaker sub-space matrix and $\boldsymbol{\epsilon}_r$ is the residual assumed to have distribution $\mathcal{N}(\textbf{0},{\boldsymbol \Sigma})$. 

For scoring, a pair of x-vectors, one from the enrollment recording ${\boldsymbol \eta}_e$ and one from the test recording ${\boldsymbol \eta}_t$ are used with the pre-trained PLDA model to compute the log-likelihood ratio score as,
\begin{equation}\label{eq:plda_scoring}
s({\boldsymbol \eta}_e, {\boldsymbol \eta}_t) = {\boldsymbol \eta}_e\transpose {\boldsymbol Q}{\boldsymbol \eta}_e + {\boldsymbol \eta}_t\transpose {\boldsymbol Q}{\boldsymbol \eta}_t + {\boldsymbol \eta}_e\transpose {\boldsymbol P}{\boldsymbol \eta}_t  
\end{equation}
where, 

\begin{eqnarray}
\textbf{Q} & = & {\boldsymbol \Sigma} _{tot} ^{-1} -  ({\boldsymbol \Sigma} _{tot} - {\boldsymbol \Sigma} _{ac} {\boldsymbol \Sigma} _{tot}^{-1} {\boldsymbol \Sigma} _{ac})^{-1} \\
\textbf{P} &  = &  {\boldsymbol \Sigma} _{tot} ^{-1} {\boldsymbol \Sigma} _{ac} ({\boldsymbol \Sigma} _{tot} - {\boldsymbol \Sigma} _{ac} {\boldsymbol \Sigma} _{tot}^{-1} {\boldsymbol \Sigma} _{ac})^{-1}
\end{eqnarray}

with ${\boldsymbol \Sigma} _{tot} = \Phi \Phi \transpose + {\boldsymbol \Sigma}$ and ${\boldsymbol \Sigma} _{ac} = \Phi \Phi \transpose $. 

\subsection{Discriminative PLDA (DPLDA)}
The discriminative PLDA proposed in  \cite{burget2011discriminatively} uses the  expanded vector $\boldsymbol{\varphi}({\boldsymbol \eta}_e,{\boldsymbol \eta}_t)$ representing a trial $({\boldsymbol \eta}_e,{\boldsymbol \eta}_t)$ was computed using a Quadratic kernel as follows: 
\begin{align}
\boldsymbol{\varphi}({\boldsymbol \eta}_e,{\boldsymbol \eta}_t) = \begin{bmatrix}
    \text{vec}({\boldsymbol \eta}_e{\boldsymbol \eta}_t\transpose + {\boldsymbol \eta}_t{\boldsymbol \eta}_e\transpose)\\
    \text{vec}({\boldsymbol \eta}_e{\boldsymbol \eta}_e\transpose + {\boldsymbol \eta}_t{\boldsymbol \eta}_t\transpose)\\
    \text{vec}({\boldsymbol \eta}_e + {\boldsymbol \eta}_t)\\
    1
    \end{bmatrix}
\end{align}
The PLDA log likelihood ratio score can be written as the dot product of a weight vector $\boldsymbol{w}$ and the expanded vector $\boldsymbol{\varphi}({\boldsymbol \eta}_e,{\boldsymbol \eta}_t)$.
\begin{equation}
    s = \boldsymbol{w}\transpose \boldsymbol{\varphi}({\boldsymbol \eta}_e,{\boldsymbol \eta}_t)
\end{equation}
Once the weight vector ${\boldsymbol w}$ is trained, the score on the test trials is generated as the inner product of the weight vector with the Quadratic kernel.

\subsection{Pairwise Gaussian backend (GB)}
The pairwise Gaussian backend \cite{ramoji2019leap,cumani2014generative}  models the pairs of enrollment and test x-vectors, ${\boldsymbol \eta} = [ {\boldsymbol \eta}_{e}\transpose \,\, {\boldsymbol \eta}_{t}\transpose ]\transpose$. The x-vector pairs are modeled using a Gaussian distribution  with parameters $(\Mu_t, {\boldsymbol \Sigma}_t)$ for target trials while the non-target pairs are modeled by a Gaussian distribution with parameters $(\Mu_{nt}, {\boldsymbol \Sigma}_{nt})$. These parameters are estimated by computing the sample mean and covariance matrices of the target and non-target trials in the training data. The log-likelihood ratio ($L$) for a new trial is then obtained as:
\begin{eqnarray}
 L = ({\boldsymbol \eta}-\Mu_{nt})\transpose{\boldsymbol \Sigma}_{nt}\inv({\boldsymbol \eta}-\Mu_{nt})-({\boldsymbol \eta}-\Mu_t)\transpose{\boldsymbol \Sigma}_t\inv({\boldsymbol \eta}-\Mu_t)  \nonumber
\end{eqnarray}

\section{Proposed Backend  - Neural PLDA}\label{sec:PldaNet}
In the proposed pairwise discriminative network (NPLDA)  (Fig.~\ref{fig:PldaNet}), we construct the pre-processing steps of LDA as first affine layer, unit-length normalization as a non-linear activation and PLDA centering and diagonalization as another affine transformation. The final PLDA pair-wise scoring given in eq.~\ref{eq:plda_scoring} is implemented as a Quadratic layer in Fig.~\ref{fig:PldaNet}. Thus, the NPLDA implements the pre-processing of the x-vectors and the PLDA scoring as a neural backend. The parameters of the NPLDA model are initialized with the baseline system and these parameters are learned in a backpropagation setting.

\subsection{Cost Function and Regularization}\label{sec:costfuncs}
To train the Neural PLDA for the task of speaker verification, we sample pairs of x-vectors representing target (from same speaker) and non-target hypothesis (from different speakers).  We train the model using randomly sampled target and non-target trials (pairs of VoxCeleb segments) which are gender matched.

In the following subsections, we describe the  common approach to error functions in the speaker verification framework which is based on binary cross entropy as well as the proposed approach based on approximating the detection cost function. 
\subsection{Binary Cross Entropy}
The standard objective for a two class classification task.
\begin{align}
    L_{BCE} = \frac{1}{N}\sum_{i=1}^{N} t_i \log {\boldsymbol \Sigma}(s_i) + (1-t_i) \log (1-{\boldsymbol \Sigma} (s_i))
\end{align}
where $s_i$ is the score for the $i$th trial, $t_i$ is the binary target for the trial and $N$ is the number of trials. 

Using this loss alone for training may result in over-fitting. Hence, a regularization term can be  used by regressing to raw PLDA scores. The regularized cross-entropy loss is given as:
\begin{align}
    L_{BCE}^{\prime} = L_{BCE} + \frac{\lambda}{N}\sum_{i=1}^{N} (s_i -l_i)^2
\end{align}
The second term encourages the scores from the Neural PLDA to not digress from the GPLDA scores drastically.

\subsection{Soft Detection Cost}
The normalized detection cost function (DCF) \cite{van2007introduction} is defined as:
\begin{align}\label{eq:det_cost}
    C_{Norm}(\beta,\theta) = P_{Miss}(\theta) + \beta P_{FA}(\theta)
\end{align}
where $\beta$ is an application based weight defined as
\begin{align}
    \beta = \frac{C_{FA} (1-P_{target})}{C_{Miss}P_{target}}
\end{align}
where $C_{Miss}$ and $C_{FA}$ are the costs assigned to miss and false alarms, and $P_{target}$ is the prior probability of a target trial.
$P_{Miss}$ and $P_{FA}$ are the probability of miss and false alarms respectively, and are computed by applying a detection threshold of $\theta$ to the log-likelihood ratios.
\begin{align}
    P_{Miss}(\theta) &= \frac{\sum_{i=1}^{N} t_i \mathds{1}(s_i<\theta)}{\sum_{i=1}^{N} t_i}\\
    P_{FA}(\theta) &= \frac{\sum_{i=1}^{N} (1-t_i) \mathds{1}(s_i \geq \theta)}{\sum_{i=1}^{N} (1-t_i)}.
\end{align}
Here, $s_i$ is the score (LLR) output by the model, $t_i$ is the ground truth variable for trial $i$. That is, $t_i = 0$ if trial $i$ is a target trial, and $t_i = 1$ if it is a non-target trial. $\mathds{1}$ is the indicator function. The normalized detection cost function (eq.~\ref{eq:det_cost}) is not a smooth function of the parameters due to the step discontinuity induced by the indicator function $\mathds{1}$, and hence, it cannot be used as an objective function in a neural network. We propose a differentiable approximation of the normalized detection cost by approximating the indicator function with a sigmoid function. 
\begin{align}
    P_{Miss}^{\text{(soft)}}(\theta) &= \frac{\sum_{i=1}^{N} t_i  \left[1-{\boldsymbol \Sigma}(\alpha(s_i-\theta))\right]}{\sum_{i=1}^{N} t_i} \\
    P_{FA}^{\text{(soft)}}(\theta) &= \frac{\sum_{i=1}^{N} (1-t_i) {\boldsymbol \Sigma}(\alpha(s_i - \theta))}{\sum_{i=1}^{N} (1-t_i)}
\end{align}
By choosing a large enough value for the warping factor $\alpha$, the approximation can be made arbitrarily close to the actual detection cost function for a wide range of thresholds.

The primary cost metric of the VOiCES challenge is the normalized detection cost function (actDCF) computed at the threshold of $\log \beta$ applied to the LLRs. is given by
\begin{align}
    C_{Primary} = C_{Norm}(\beta), \log \beta_e)
\end{align}
where $\beta = 99$. We compute the Neural PLDA loss function as
\begin{align}\label{eq:softdcf}
     \mathcal{L}_{Primary} = C_{Norm}^{\text{(soft)}}(\beta, \theta)
\end{align}
where $\theta$ is the thresholds which minimizes $\mathcal{L}_{Primary}$ when included as a model parameter. The minimum detection cost (minDCF) is achieved at a threshold where the DCF is minimized.
\begin{align}
    \text{minDCF} = \underset{\theta}{\min} \,\,C_{Norm}(\beta, \theta)
\end{align}
In other words, it is the best cost that can be achieved through calibration of the scores. We include these thresholds in the set of parameters that the neural network learns to minimize minDCF through backpropagation.

 \begin{figure}[t]
 \begin{center}
    \includegraphics[width=\linewidth, trim={0cm 0.1cm 1cm 1cm},clip]{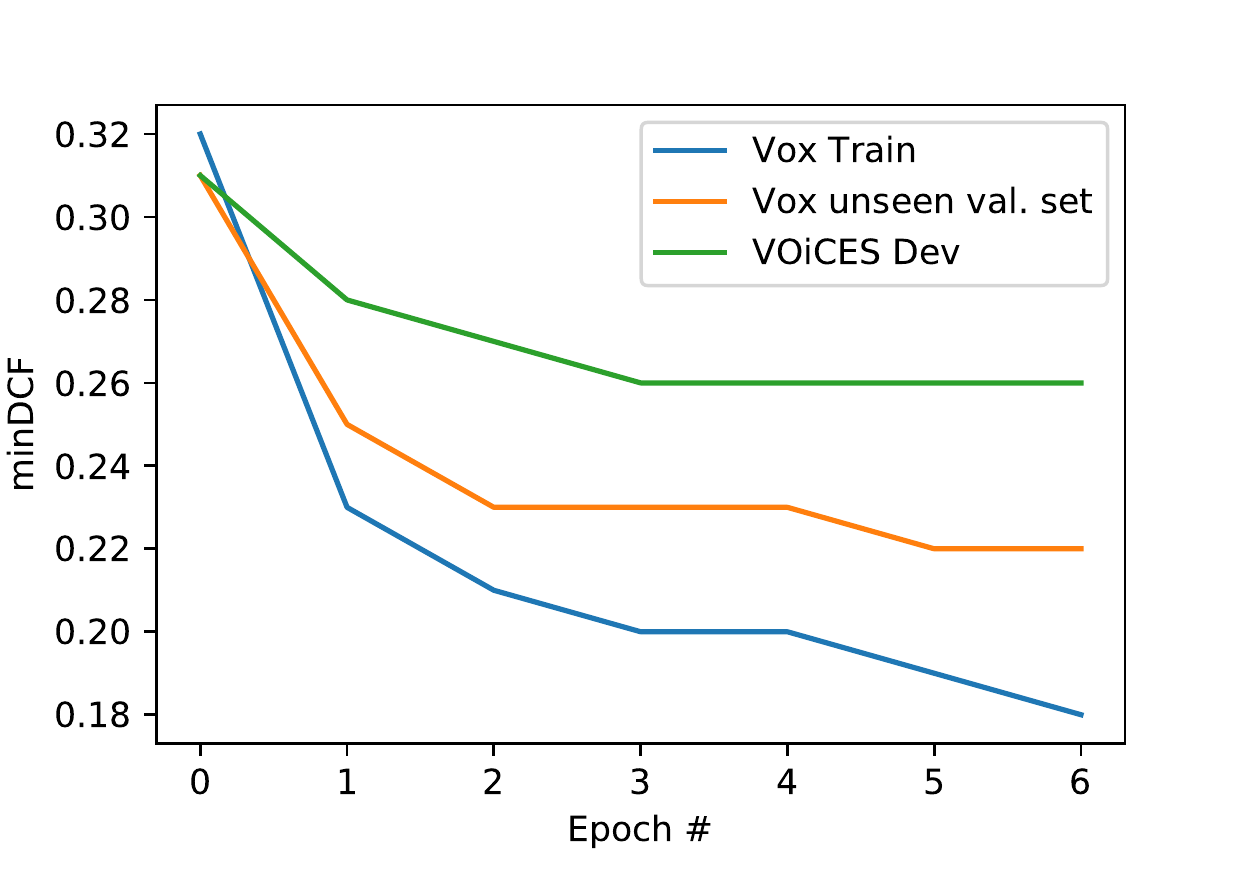}
    \vspace{-0.1cm}
    \caption{Plot of minDCF as function of the  epochs in the NPLDA training. The best model in Vox validation set is used as the final model. Note that the minDCF on iteration $0$ corresponds to the generative PLDA model. }
    \label{fig:training_plot}
\end{center}
\vspace{-0.1cm}
\end{figure}

\subsection{Model Training}
We performed various experiments using the NPLDA architecture with different initialization methods and loss functions. We also experiment with the role of batch size parameter, the learning rate as well as the choice of loss function in the optimization. The optimal parameter choices were based on the VOiCES development set. In this work, we also parameterize the threshold value and let the network learn the threshold value to minimize the loss (Eq.~\ref{eq:softdcf}).

The soft detection cost function is highly sensitive to small changes in false alarm probability. As any imbalance in target to non-target trial ratio in the mini-batches impacts the NPLDA model training,  we choose a large batch size for training the NPLDA  network ($8192$). The learning rate was initialized to $10^{-3}$ and halved each time the validation loss failed to decrease for two epochs.
We use the Adam optimizer for the backpropagation learning and the entire model is implemented in the PyTorch\blfootnote{The implementation of the NPLDA Network can be found here: \url{https://github.com/iiscleap/NeuralPlda}} toolkit~\cite{paszke2019pytorch}.

In Fig.~\ref{fig:training_plot}, we show the variation of the minimum detection cost fuction (minDCF), as function of the training epochs. The model parameters are initialized using the generative PLDA model backend. Specifically, the NPLDA parameters like the LDA, WCCN, the diagonalizing transforms and the $\textbf{P},\textbf{Q}$ matrices used in the scoring are initialized using the baseline model. The results on the training and validation trials for epoch $0$ corresponds to this baseline model. As seen in this plot, the backpropagation of the proposed model loss function significantly improves the minDCF for the training, validation as well as the VOiCES development set. The best performance in the validation dataset (Vox validation) is achieved at epoch 6,which is used for the final evaluation.

\begin{table*}[t]
\centering
\label{tab:results}
\begin{tabular}{@{}l|l|cc|cc|cc@{}}
\toprule
\multirow{2}{*}{Model} & \multirow{2}{*}{PLDA Train Dataset} & \multicolumn{2}{c|}{SITW Core-Core} & \multicolumn{2}{c|}{VOiCES Dev} & \multicolumn{2}{c}{VOiCES Eval} \\ \cmidrule(l){3-8} 
                       &                                 & EER (\%)           & minDCF           & EER (\%)           & minDCF           & EER (\%)      & minDCF        \\ \midrule
GPLDA   &  VoxCeleb                               & 2.79      & 0.29                & 2.79                & 0.31               &  7.35        & 0.57        \\
GPLDA                      &   VoxCeleb Augmented                          & 2.79               & 0.29               & 2.79               &     0.30           & 6.38         &  0.53                   \\
 Gaussian Backend                       & VoxCeleb                                 & 3.19              &   0.31             & 3.14                & 0.33               & 7.58         & 0.60                    \\
  Gaussian Backend                       & VoxCeleb Augmented                                 & 3.06              &   0.31             &    2.89             & 0.30               & 6.63         & 0.53                    \\
DPLDA   & VoxCeleb Augmented  & 2.98       &  0.32     & 3.05    & 0.36  & 6.65 & 0.56 \\
NPLDA                       &     VoxCeleb                            & 2.14              &  0.23              & 2.20                & 0.26               & 6.72         & 0.51                    \\
NPLDA                       & VoxCeleb    Augmented                            &  \textbf{2.05}             & \textbf{0.20}               & \textbf{1.91}                & \textbf{0.23 }           & \textbf{6.01}          &    \textbf{0.49}                 \\
NPLDA (BCE Loss)  & VoxCeleb Augmented  &  2.10      &   0.22    &   2.32  & 0.26  & 6.34 & 0.53 \\
\bottomrule
\end{tabular}
\caption{Performance of systems on SITW Eval Core-Core, VOiCES Dev and VOiCES Eval using the GPLDA baseline model, Gaussian backend, Discriminative PLDA (DPLDA) and the proposed NPLDA model. We also report the use of binary cross-entropy (BCE) Loss in the NPLDA model place of the soft detection cost. The best scores are highlighted.}
\end{table*}

\section{Experiments}\label{sec:results}
We perform several experiments with the proposed neural net architecture and compare them with  various discriminative backends previously proposed in the literature such as the discriminative PLDA \cite{burget2011discriminatively, glembek2011discriminatively} and pairwise Gaussian backend~\cite{cumani2013pairwise}. We also compare the performance with the baseline system using Kaldi recipe that implements the generative PLDA model based scoring.

For all the pairwise generative/discriminative models, we train the backend using randomly sampled target and non-target pairs which are matched by gender. We perform experiments by sampling trials from the clean VoxCeleb segments, and also the augmented set. We sample about $6.6$ million trials from the clean set and around $33$ million trials from the augmented set. 

\subsection{GPLDA Baseline}
The primary baseline to benchmark our systems is the PLDA backend implementation in the Kaldi toolkit. The Kaldi implementation models the average embedding x-vector of each training speaker. The x-vectors are centered, dimensionality reduced using LDA to 170 dimensions, followed by unit length normalization. The linear transformations and the GPLDA matrices are used to initialize the proposed pairwise PLDA network.

\subsection{Gaussian Backend}
The Gaussian backend is also trained on the same pairs of target and non-target x-vector trials, after centering, LDA and length normalization. The procedure to sample the trials is similar to the one used in our previous work \cite{ramoji2019leap}. We randomly sample pairs of gender matched x-vectors from Voxceleb training dataset belonging to target and non-target trials. This generates a total of $5$M trials for the Gaussian Backend training. A ratio of $1$:$10$ is used for the target to non-target ratio in the training trials.

\subsection{Discriminative PLDA (DPLDA)}
The procedure to sample trials is similar to what we used for the pairwise Gaussian Backend model. A portion of the Voxceleb training trials is held out as validation (unseen speakers). We also use a trial ratio of $1$:$10$ in the training and validation set for target/non-target ratio. The x-vectors are centered, LDA transformed and length normalized the same way as in the GPLDA baseline. A discriminative PLDA model \cite{burget2011discriminatively} is trained on these processed x-vectors.

\subsection{Neural PLDA (NPLDA)}
The same set of trials used in DPLDA are used to train the NPLDA model. We initialize the model with a pretrained GPLDA from Kaldi, and optimize the NPLDA network with the soft detection cost. We also report the performance of this architecture with the same initialization, using the binary cross-entropy (BCE) to provide comparisons with the soft detection cost function.

\subsection {Discussion of Results}
The performance of the various backend systems are reported in Table~\ref{tab:results}. The GPLDA baseline trained using the Voxceleb dataset shows comparable performance for the SITW and the VOiCES Development dataset. However, the VOiCES evaluation dataset has a significant degradation in the performance for the GPLDA model. The data augmentation of the Voxceleb dataset improves the GPLDA model performance for the VOiCES evaluation dataset. The Gaussian Backend is marginally worse than the GPLDA model for all the test conditions. The DPLDA model was found to be moderately worse than the GPLDA model in all the test conditions where the Voxceleb augmented set was used for PLDA training. 

The NPLDA model improves significantly over the GPLDA model in all the test conditions.  Without any augmentation, the model improves over the baseline GPLDA system in terms of EER (relative improvements of $23$\%, $22$\% and $9$ \% for SITW, VOiCES Dev. and VOiCES Eval respectively) as well as in terms of minDCF (relative improvements of $31$\%, $20$\% and $11$ \% for SITW, VOiCES Dev. and VOiCES Eval respectively). These improvements are consistent with data augmentation as well. The results highlight that NPLDA based optimization combines many sub-modules like discriminant analysis, covariance normalization as well as PLDA scoring into a single efficient pipeline.

When the binary cross-entropy (BCE) loss is used in place of the soft detection cost, the performance improves over the GPLDA baseline for the SITW core-core and VOiCES dev set, but are found to be worse than its counterpart which uses the soft detection cost. However, for the VOiCES evaluation set, the performance is only marginally better than the GPLDA (in terms of EER) and same in terms of minDCF upto two decimal places. This clearly indicates that the soft detection cost is a better choice of cost function for the NPLDA model.

\section{Summary and Conclusions}\label{sec:summary}
This paper presents a step in the direction of exploring discriminative models for the task of speaker verification. Discriminative models allow the construction of end-to-end systems. However, discriminative models tend to overfit to the training data. In our proposed model, we constrain the parameter set to have lesser degrees of freedom, in order to achieve better generalization. We also propose a task specific differentiable loss function which approximates the detection cost function. 

Using a single elegant backend model that is targeted to optimize the speaker verification loss, the NPLDA model uses the x-vector embeddings directly to generate the speaker verification score. The experiments on the VOiCES and SITW datasets illustrate the performance gains obtained for the proposed approach. 

In the proposed approach, the  fully neural backend and a neural embedding extractor are used. The future research work will explore the combination of the frontend and the backend into a single neural end-to-end engine for speaker verification.

\section{Acknowledgements}
This work was funded by the Ministry of Human Resources Development (MHRD) of India and the Department of Science and Technology (DST) EMR/2016/007934 grant.

\bibliographystyle{IEEEbib}
\bibliography{Odyssey2020_Voices}

\begin{thebibliography}{10}

\bibitem{reynolds2000speaker}
Douglas~A Reynolds, Thomas~F Quatieri, and Robert~B Dunn,
\newblock ``Speaker verification using adapted gaussian mixture models,''
\newblock {\em Digital signal processing}, vol. 10, no. 1-3, pp. 19--41, 2000.

\bibitem{kenny2007joint}
Patrick Kenny, Gilles Boulianne, Pierre Ouellet, and Pierre Dumouchel,
\newblock ``Joint factor analysis versus eigenchannels in speaker
  recognition,''
\newblock {\em IEEE Transactions on Audio, Speech, and Language Processing},
  vol. 15, no. 4, pp. 1435--1447, 2007.

\bibitem{dehak2011front}
Najim Dehak, Patrick~J Kenny, R{\'e}da Dehak, Pierre Dumouchel, and Pierre
  Ouellet,
\newblock ``Front-end factor analysis for speaker verification,''
\newblock {\em IEEE Transactions on Audio, Speech, and Language Processing},
  vol. 19, no. 4, pp. 788--798, 2011.

\bibitem{snyder2018x}
David Snyder, Daniel Garcia-Romero, Gregory Sell, Daniel Povey, and Sanjeev
  Khudanpur,
\newblock ``X-vectors: Robust dnn embeddings for speaker recognition,''
\newblock in {\em ICASSP 2018}. IEEE, 2018, pp. 5329--5333.

\bibitem{mclaren2018train}
Mitchell Mclaren, Diego Cast{\'a}n, Mahesh~Kumar Nandwana, Luciana Ferrer, and
  Emre Yilmaz,
\newblock ``How to train your speaker embeddings extractor,''
\newblock in {\em Proc. Odyssey 2018 The Speaker and Language Recognition
  Workshop}, 2018, pp. 327--334.

\bibitem{garcia2011analysis}
Daniel Garcia-Romero and Carol~Y Espy-Wilson,
\newblock ``Analysis of i-vector length normalization in speaker recognition
  systems,''
\newblock in {\em Interspeech 2011}, 2011.

\bibitem{hatch2006within}
Andrew~O Hatch, Sachin Kajarekar, and Andreas Stolcke,
\newblock ``Within-class covariance normalization for svm-based speaker
  recognition,''
\newblock in {\em Ninth international conference on spoken language
  processing}, 2006.

\bibitem{kenny2010bayesian}
Patrick Kenny,
\newblock ``Bayesian speaker verification with heavy-tailed priors.,''
\newblock in {\em Odyssey}, 2010, pp. 14--21.

\bibitem{van2007introduction}
David~A Van~Leeuwen and Niko Br{\"u}mmer,
\newblock ``An introduction to application-independent evaluation of speaker
  recognition systems,''
\newblock in {\em Speaker classification I}, pp. 330--353. Springer, 2007.

\bibitem{campbell2006support}
William~M Campbell, Douglas~E Sturim, and Douglas~A Reynolds,
\newblock ``Support vector machines using {GMM} supervectors for speaker
  verification,''
\newblock {\em IEEE Signal Process. Lett.}, vol. 13, no. 5, pp. 308--311, 2006.

\bibitem{mclaren2013adaptive}
Mitchell McLaren, Aaron Lawson, Yun Lei, and Nicolas Scheffer,
\newblock ``Adaptive {G}aussian backend for robust language identification.,''
\newblock in {\em Interspeech 2013}, 2013, pp. 84--88.

\bibitem{benzeghiba2009language}
Mohamed~Faouzi BenZeghiba, Jean-Luc Gauvain, and Lori Lamel,
\newblock ``Language score calibration using adapted gaussian back-end,''
\newblock in {\em Interspeech 2009}, 2009.

\bibitem{cumani2013pairwise}
Sandro Cumani, Niko Br{\"u}mmer, Luk{\'a}{\v{s}} Burget, Pietro Laface,
  Old{\v{r}}ich Plchot, and Vasileios Vasilakakis,
\newblock ``Pairwise discriminative speaker verification in the i-vector
  space,''
\newblock {\em IEEE Transactions on Audio, Speech, and Language Processing},
  vol. 21, no. 6, pp. 1217--1227, 2013.

\bibitem{cumani2014large}
Sandro Cumani and Pietro Laface,
\newblock ``Large-scale training of pairwise support vector machines for
  speaker recognition,''
\newblock {\em IEEE/ACM Transactions on Audio, Speech and Language Processing
  (TASLP)}, vol. 22, no. 11, pp. 1590--1600, 2014.

\bibitem{cumani2014generative}
Sandro Cumani and Pietro Laface,
\newblock ``Generative pairwise models for speaker recognition,''
\newblock in {\em Odyssey}, 2014, pp. 273--279.

\bibitem{burget2011discriminatively}
Luk{\'a}{\v{s}} Burget, Old{\v{r}}ich Plchot, Sandro Cumani, Ond{\v{r}}ej
  Glembek, Pavel Mat{\v{e}}jka, and Niko Br{\"u}mmer,
\newblock ``Discriminatively trained probabilistic linear discriminant analysis
  for speaker verification,''
\newblock in {\em ICASSP 2011}. IEEE, 2011, pp. 4832--4835.

\bibitem{villalba2019state}
Jes{\'u}s Villalba, Nanxin Chen, David Snyder, Daniel Garcia-Romero, Alan
  McCree, Gregory Sell, Jonas Borgstrom, Leibny~Paola Garc{\'\i}a-Perera, Fred
  Richardson, R{\'e}da Dehak, et~al.,
\newblock ``State-of-the-art speaker recognition with neural network embeddings
  in {NIST SRE18} and speakers in the wild evaluations,''
\newblock {\em Computer Speech \& Language}, p. 101026, 2019.

\bibitem{ferrer2019optimizing}
Luciana Ferrer and Mitchell McLaren,
\newblock ``Optimizing a speaker embedding extractor through backend-driven
  regularization,''
\newblock {\em Interspeech 2019}, pp. 4350--4354, 2019.

\bibitem{rohdin2018end}
Johan Rohdin, Anna Silnova, Mireia Diez, Old{\v{r}}ch Plchot, Pavel
  Mat{\v{e}}jka, and Luk{\'a}{\v{s}} Burget,
\newblock ``End-to-end {DNN} based speaker recognition inspired by i-vector and
  {PLDA},''
\newblock in {\em ICASSP 2018}. IEEE, 2018, pp. 4874--4878.

\bibitem{zhang2017end}
Chunlei Zhang and Kazuhito Koishida,
\newblock ``End-to-end text-independent speaker verification with triplet loss
  on short utterances.,''
\newblock in {\em Interspeech 2017}, 2017, pp. 1487--1491.

\bibitem{wan2018generalized}
Li~Wan, Quan Wang, Alan Papir, and Ignacio~Lopez Moreno,
\newblock ``Generalized end-to-end loss for speaker verification,''
\newblock in {\em ICASSP 2018}. IEEE, 2018, pp. 4879--4883.

\bibitem{Mingote2019}
Victoria Mingote, Antonio Miguel, Dayana Ribas, Alfonso Ortega, and Eduardo
  Lleida,
\newblock ``{Optimization of False Acceptance/Rejection Rates and Decision
  Threshold for End-to-End Text-Dependent Speaker Verification Systems},''
\newblock in {\em Proc. Interspeech 2019}, 2019, pp. 2903--2907.

\bibitem{Nandwana2019}
Mahesh~Kumar Nandwana, Julien van Hout, Colleen Richey, Mitchell McLaren,
  Maria~A. Barrios, and Aaron Lawson,
\newblock ``{The VOiCES from a Distance Challenge 2019},''
\newblock in {\em Interspeech 2019}, 2019, pp. 2438--2442.

\bibitem{nagrani2017voxceleb}
{Nagrani, Arsha and Chung, Joon Son and Zisserman, Andrew},
\newblock ``Voxceleb: a large-scale speaker identification dataset,''
\newblock {\em arXiv preprint arXiv:1706.08612}, 2017.

\bibitem{chungvoxceleb2}
Joon~Son Chung, Arsha Nagrani, and Andrew Zisserman,
\newblock ``Voxceleb2: Deep speaker recognition,''
\newblock in {\em Interspeech 2018}, 2018, pp. 1086--1090.

\bibitem{nandwana2019voices}
Mahesh~Kumar Nandwana, Julien Van~Hout, Mitchell McLaren, Colleen Richey, Aaron
  Lawson, and Maria~Alejandra Barrios,
\newblock ``The voices from a distance challenge 2019 evaluation plan,''
\newblock {\em arXiv preprint arXiv:1902.10828}, 2019.

\bibitem{McLaren+2016}
Mitchell McLaren, Luciana Ferrer, Diego Castan, and Aaron Lawson,
\newblock ``The speakers in the wild (sitw) speaker recognition database,''
\newblock in {\em Interspeech 2016}, 2016, pp. 818--822.

\bibitem{sizov2014unifying}
Aleksandr Sizov, Kong~Aik Lee, and Tomi Kinnunen,
\newblock ``Unifying probabilistic linear discriminant analysis variants in
  biometric authentication,''
\newblock in {\em Joint IAPR International Workshops on Statistical Techniques
  in Pattern Recognition (SPR) and Structural and Syntactic Pattern Recognition
  (SSPR)}. Springer, 2014, pp. 464--475.

\bibitem{ramoji2019leap}
{Ramoji, Shreyas and Mohan, Anand and Mysore, Bhargavram and Bhatia, Anmol and
  Singh, Prachi and Vardhan, Harsha and Ganapathy, Sriram},
\newblock ``{The Leap Speaker Recognition System for NIST SRE 2018
  Challenge},''
\newblock in {\em ICASSP 2019}. IEEE, 2019, pp. 5771--5775.

\bibitem{paszke2019pytorch}
Adam Paszke, Sam Gross, Francisco Massa, Adam Lerer, James Bradbury, Gregory
  Chanan, Trevor Killeen, Zeming Lin, Natalia Gimelshein, Luca Antiga, et~al.,
\newblock ``Pytorch: An imperative style, high-performance deep learning
  library,''
\newblock in {\em Advances in Neural Information Processing Systems}, 2019.

\bibitem{glembek2011discriminatively}
Ond{\v{r}}ej Glembek, Luk{\'a}{\v{s}} Burget, Niko Br{\"u}mmer, Old{\v{r}}ich
  Plchot, and Pavel Mat{\v{e}}jka,
\newblock ``Discriminatively trained i-vector extractor for speaker
  verification,''
\newblock in {\em Interspeech 2011}, 2011.

\end{thebibliography}

%

\end{document}